\documentclass[12pt]{iopart}

\usepackage{graphicx}                  % Include figure files
\usepackage{dcolumn}                   % Align table columns on decimal point
\usepackage{bm}                        % bold math

\usepackage{color}

\newcommand {\E}  {{\varepsilon}}

\newcommand{\lamu}{{\lambda_{\rm u}}}
\newcommand{\Omu}{{\Omega_{\rm u}}}

\newcommand{\Omch}{{\Omega_{\rm ch}}}

\begin{document}

\title{Radiation from multi-GeV electrons and positrons in periodically bent silicon crystal}

\author{
Victor~G.~Bezchastnov$^1$
\footnote{On leave from A.F.~Ioffe Physical-Technical Institute, 
Politechnicheskaya Str. 26, 194021 St. Petersburg, Russia and St. Petersburg 
State Polytechnical University, Politechnicheskaya 29, 195251 St. Petersburg, Russia}, 
Andrei~V.~Korol$^2$
\footnote{Permanent address: St. Petersburg State Maritime University, 
Leninsky ave. 101, 198262 St. Petersburg, Russia}
and Andrey~V.~Solov'yov$^2$
\footnote{On leave from A.F.~Ioffe Physical-Technical 
Institute, Politechnicheskaya Str. 26, 194021 St. Petersburg, Russia}
}

\address{$^1$ Institute for Nuclear Physics, Johannes Gutenberg University, 
Johann-Joachim-Becher-Weg~45, D-55128 Mainz, Germany}

\address{$^2$ MBN Research Center, Department of Physics, Goethe University, 
Altenh{\"o}ferallee~3, D-60438 Frankfurt am Main, Germany}

%\date{\today}

\begin{abstract}
A periodically bent Si crystal is shown to efficiently serve for producing highly monochromatic 
radiation in a gamma-ray energy spectral range. 
A short-period small-amplitude bending yields narrow undulator-type spectral 
peaks in radiation from multi-GeV electrons and positrons channeling through the crystal. 
Benchmark theoretical results on the undulator are obtained by simulations of the channeling 
with a full atomistic approach to the projectile-crystal interactions over the 
macroscopic propagation distances. 
The simulations are facilitated by employing the MBN Explorer package 
for molecular dynamics calculations on the meso- bio- and nano-scales. 
The radiation from the ultra-relativistic channeling projectiles is computed within the 
quasi-classical formalism. The effects due to the quantum recoil are shown to be significantly 
prominent in the gamma-ray undulator radiation. 
\end{abstract}


\maketitle

The phenomena of channeling of relativistic charged particles through the crystalline structures 
remain in focus of challenging studies, both theoretical and experimental ones, for a number of 
years (see, e.g., the recent monograph~\cite{ChannelingBook2013} and the references therein). 
In the channeling mode, the particles move along crystallographic planes or axes such that the 
trajectories exhibit bound oscillations. Out of this mode, the motion is entirely unbound. 
The two types of motion are very contrast in properties, and so are the radiations produced by 
these motions. The radiation from channeling particles possesses distinct spectral features 
signifying bound oscillations in the trajectories, in contrast to a broad-spectrum incoherent 
bremsstrahlung radiation from the non-channeling projectiles.

A variety of conditions have been considered in theoretical simulations on channeling dynamics 
and radiation, to be mentioned among the many are the recent 
studies~\cite{NewPaper_2013,JPCS_2013a,JPCS_2013b,K_PRL_2013,Mainz_PRL_2014}. 
Most widely addressed is the planar channeling in straight crystals. 
Here, the bound-type motions of the projectiles are the {\em channeling oscillations} in the 
direction transverse to the planes. The corresponding frequencies $\Omch$ are 
determined by the crystalline inter-planar potentials and by the projectile energies. 
In periodically bent crystals, the transverse motions of the projectiles acquire additional 
{\em undulator oscillations}. 
For relativistic particles, the velocities of motion along the channel are close to the light 
speed $c$, and the undulator oscillation frequency is $\Omu = 2\pi{c}/\lamu$, where $\lamu$ 
is the bending period of the crystal. 

The oscillations in motion yield the spectral lines in the radiation emitted by the projectiles. 
For the ultra-relativistic projectiles, the radiation emerges in a narrow domain of the directions 
close to the direction of the mean velocity, and the line positions scale according 
to the second power of the Lorentz-factor $\gamma$ of the projectiles~\cite{Jackson}. 
Possibility to tune these positions by varying the bending period and the projectile energies 
is an attractive property of the crystalline undulator (CU), a device whose concept was 
formulated some time ago~\cite{KSG1998,KSG_review_1999}. Theoretically, it has been rationalized 
that a CU can produce the monochromatic radiation of sub-MeV to MeV energies, and different 
schemes for CUs have been proposed for experiments. An evident experimental manifestation 
of the CU radiation has yet to happen, and further theoretical studies are important for 
supporting the experiments.

Since introducing the concept of CU, major theoretical studies on stability of the undulator 
motions as well as the calculations of the emerging radiation have been concerning the regimes 
where the particles follow the periodically bent crystalline shapes. For such motions, 
the undulator oscillation frequencies are smaller than the frequencies of the channeling 
oscillations, and hence the undulator lines appear in the spectra at the energies below the 
energies of the channeling lines. Yet another regime, with the opposite relation  
$\Omu > \Omch$ between the oscillation frequencies, has recently been shown~\cite{K_PRL_2013} 
possible. There, the particles do not follow the bent profiles and the channeling 
resembles that in the straight crystal. The latter regime can be achieved in a 
{\em short-period small-amplitude} CU where the bending period $\lamu$ is shorter than the 
period of channeling oscillations in the straight crystal whereas the bending amplitude $a$ 
is smaller than the inter-planner distance $d$. 

The newly proposed short-period small-amplitude CU is a novel one and certainly 
requires additional theoretical verifications in order to be put forward for the experiments. 
The simulations reported in Ref.~\cite{K_PRL_2013} are for the $855$~MeV electrons and positrons 
and come along with the conclusion that the effects of bending are not seen in the trajectories 
but displayed in the radiation spectra. Such a puzzling outcome has to be verified in order to 
elucidate the physics of the undulator radiation. This has motivated us to investigate the 
subject by the theoretical methods more advanced and accurate that those employed in the 
studies~\cite{K_PRL_2013}. 

It is also demanding to study impact of the short-period small-amplitude bending on 
the electrons and positrons with energies higher than the value $855$~MeV 
considered in Ref.~\cite{K_PRL_2013}. Current experiments with CUs made of 
Si(110) extend to the multi-GeV projectile energies, in particular to the value of 
$10$~GeV provided by the SLAC experimental facility~\cite{Ugg}. 
With increasing the beam energy, the undulator radiation, if produced, appears in a harder 
spectral domain and becomes more distinct in energies from a much softer channeling radiation. 
The increasing energy separation between the channeling and undulator lines is very attractive 
for the experimental detection of the undulator effect and its applications. 
We therefore focus on the $10$~GeV projectiles, for the set of CU parameters 
considered in~\cite{K_PRL_2013}. 

To address the essence of undulator radiation it is imperative to study whether the projectiles 
acquire the short-period bound oscillations when passing through the CU. Giving the positive 
answer leaves no puzzle in understanding the spectral lines in CU radiation. 
For the multi-GeV particles and the bending periods of hundreds of nanometers, 
these lines appear at gamma-ray energies only a few times smaller than the energy 
of the projectiles. Here, the {\em quantum effects} in the CU radiation turn out to be important, 
in particular these due to the {\em quantum recoil}, which were neglected in~\cite{K_PRL_2013} 
for the sub-GeV projectiles. Along with providing reliable simulations on the CU radiation, 
it is demanding to uncover the significance of the recoil in the spectra, the issue 
which to the best of our knowledge has yet not been quantitatively examined. 

To simulate the channeling dynamics, we employ, as in the previously reported 
studies~\cite{JPCS_2013a,JPCS_2013b}, 
the MBN Explorer package~\cite{MBN_ExplorerPaper,MBN_ExplorerSite}. This package is designed 
for high-accuracy computations on a broad variety of phenomena developing at  
meso- bio- and nano-scales in space and time. The theoretical approaches 
to channeling simulations with the MBN Explorer advance beyond the approximations applied 
in the other studies, in particular with respect to computing the projectile-crystal 
interaction. The case of multi-GeV projectiles is also an important 
benchmark to verify the capabilities of the MBN explorer. 

To compute the radiation from the channeling motion, we use the quasi-classical formalism 
developed by Baier~{\em et al}~\cite{Baier_1988}. Here, the radiative transitions of the 
projectiles are described in terms of the classical trajectories, which allows for a direct link 
of the calculations to the results on the channeling dynamics. 
The quasi-classical formalism neglects the quantization of the projectile's motion, which is a 
safe approximation for the energies of the interest. At the same time, 
this approach includes the effects due to the quantum recoil important for the photon 
emission energies $\omega$ being no longer negligible compared to the energies $\E$ of the projectiles. 

The spectral intensities are computed by numerical integrations over the trajectories, as described 
in detail in Ref.~\cite{NewPaper_2013}. For the periodic motions, a variety of analytical 
approximations become possible. In particular, the following relation~\cite{ChannelingBook2013} 
applies to the positions of the undulator peaks in the radiation along the mean velocity of 
the projectiles:
\begin{equation}
\omega'_n = \frac{\omega_n}{1-(\omega_n/\E)} = \frac{2\gamma^2\,\Omu\,n}{1+(K^2/2)},
\;\;\;
n = 1,2,3,\ldots.
\label{peak_positions}
\end{equation}
The integer $n$ enumerates the fundamental ($n=1$) peak and the higher ($n>1$) harmonics, with 
$\omega_n$ and $\omega'_n$ being the peak positions with account for the quantum recoil and 
neglecting the recoil, respectively. Due to the recoil, the radiation frequencies become smaller 
than the frequencies neglecting the recoil, $\omega_n < \omega'_n$. 
The undulator parameter $K$ results from the coupling 
between the projectile's motion along the mean velocity and the oscillations 
in the transverse direction. 
We use hereafter the energy units for the frequencies and notice that the ratio $\omega_n/\E$ 
is a {\em quantum parameter}. We will refer to Eq.~(\ref{peak_positions}) 
when discussing the results of calculations on the CU radiation. 

Next we briefly outline the details of the simulations on the channeling dynamics.  
As we study the channeling in Si(110), the (110) planes of a straight crystal define the 
$xz$-plane for our reference coordinate system. The $z$-axis is assigned to the direction 
of the incoming projectiles. A care is taken to avoid accidental orientations of the beam 
along crystallographic axes, as this is also done in the most common experiments in order to 
exclude the axial channeling. 
To simulate a bent crystal with the MBN Explorer, the coordinates $x',y',z'$ of 
each lattice node are obtained from the coordinates $x,y,z$ of the same node in a straight 
crystal according to the relations 
$x' = x$, $y' = y + a\cos(2\pi{z}/\lamu)$ and $z' = z$. 
For the short-period small-amplitude CUs of our interest, the values 
of $\lamu$ and $a$ are of the hundreds of nanometers and of a few tens of Angstroms, 
respectively. In order to account for the thermal vibrations, the lattice 
structure is additionally transformed by random displacements of the atoms with respect 
to the equilibrium positions. The displacements obey the normal distribution corresponding 
to the value $0.075$~{\AA} for the amplitude of the vibrations. 

To uncover a maximum outcome of a possible experiment, we simulate a zero-emittance beam 
at the crystalline entrance $z=0$. The incoming particles are directed along the (110) 
planes of the reference straight crystal, and the Monte-Carlo sampling over the 
transverse coordinates covers all different entrance points available for the beam to get 
into the inside of the crystal. With the above initial conditions, the MBN Explorer solves 
numerically the relativistic equations of motion to determine the projectile's trajectories. 
In course of the integration, the interaction of the projectiles with the crystalline 
environment is computed as a multi-center interaction with all those 
atoms that influence the motion. The latter atoms are dynamically selected within 
appropriately large simulation box, and the single interactions are described by the 
Moliere potentials. The relevant details can be found in, e.g., Ref.~\cite{NewPaper_2013}. 
Here we would like to stress that, the above approach treats the forces experienced 
by the projectiles in a much more accurate way than the commonly employed mean-field 
approach or a particular ``snap-shot'' approximation to the forces employed in 
the studies~\cite{K_PRL_2013}. With the MBN Explorer, the robust Moliere interactions 
are summed up over the macroscopic propagation sizes and the numbers of the atoms. 
For example, for the propagation length of $10~\mu$m in Si, the interactions with 
approximately $2.5 \times 10^6$ atoms are included into the numerical simulations 
on the dynamics of the projectiles. For each set of CU parameters, sampling the 
initial conditions gave rise to more than $3000$ trajectories that have been computed.  

\begin{figure}[th]
%\hspace{-3mm}
\includegraphics[width=\columnwidth]{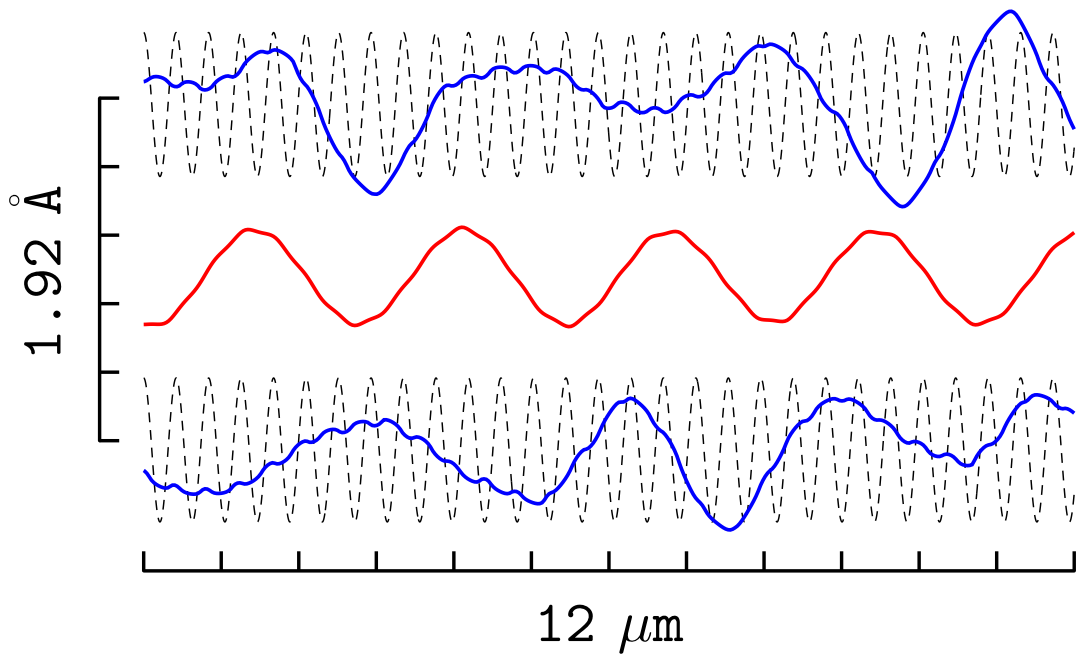}
%\vspace{-5mm}
\caption{(color online). Effect of the short-period ($\lamu = 400$~nm) small-amplitude 
($a = 0.4$~{\AA}) bending on channeling of the $855$~GeV projectiles in Si. Solid lines 
are the $12~\mu$m long trajectory segments for the channeling motion, whereas the dashed 
lines are the profiles of the bent crystalline (110) planes. 
The distance between the mid-lines of the profiles is equal to the inter-planner 
distance in the straight crystal ($1.92$~{\AA}) and is marked by the vertical scale. 
The electron and positron trajectories spread through and in between the bent profiles 
and are colored blue and red, respectively, in the color figure. 
Clearly seen in the trajectories are short-period modulations resulting from the bent 
crystalline structure.}
\label{trajectories_855}
%\vspace{-2mm}
\end{figure}

First we have simulated propagation of the $855$~MeV projectiles in Si CU with 
$\lamu = 400$~nm and $a = 0.4$~{\AA}. Opposite to the conclusion of Ref.~\cite{K_PRL_2013} 
for this CU and the beam energy, we find the effect of bending to be well displayed in the 
trajectories of the projectiles. This is demonstrated by Figure~\ref{trajectories_855} 
showing the typical segments of the channeling motion over the $12~\mu$m propagation 
scale along the bent crystal. The small yet well seen oscillations in the trajectories 
are evident to acquire the undulator period $\lamu$ and thus to be responsible for the 
undulator lines in the radiation spectra. 

\begin{figure}[th]
%\hspace{-3mm}
\includegraphics[width=\columnwidth]{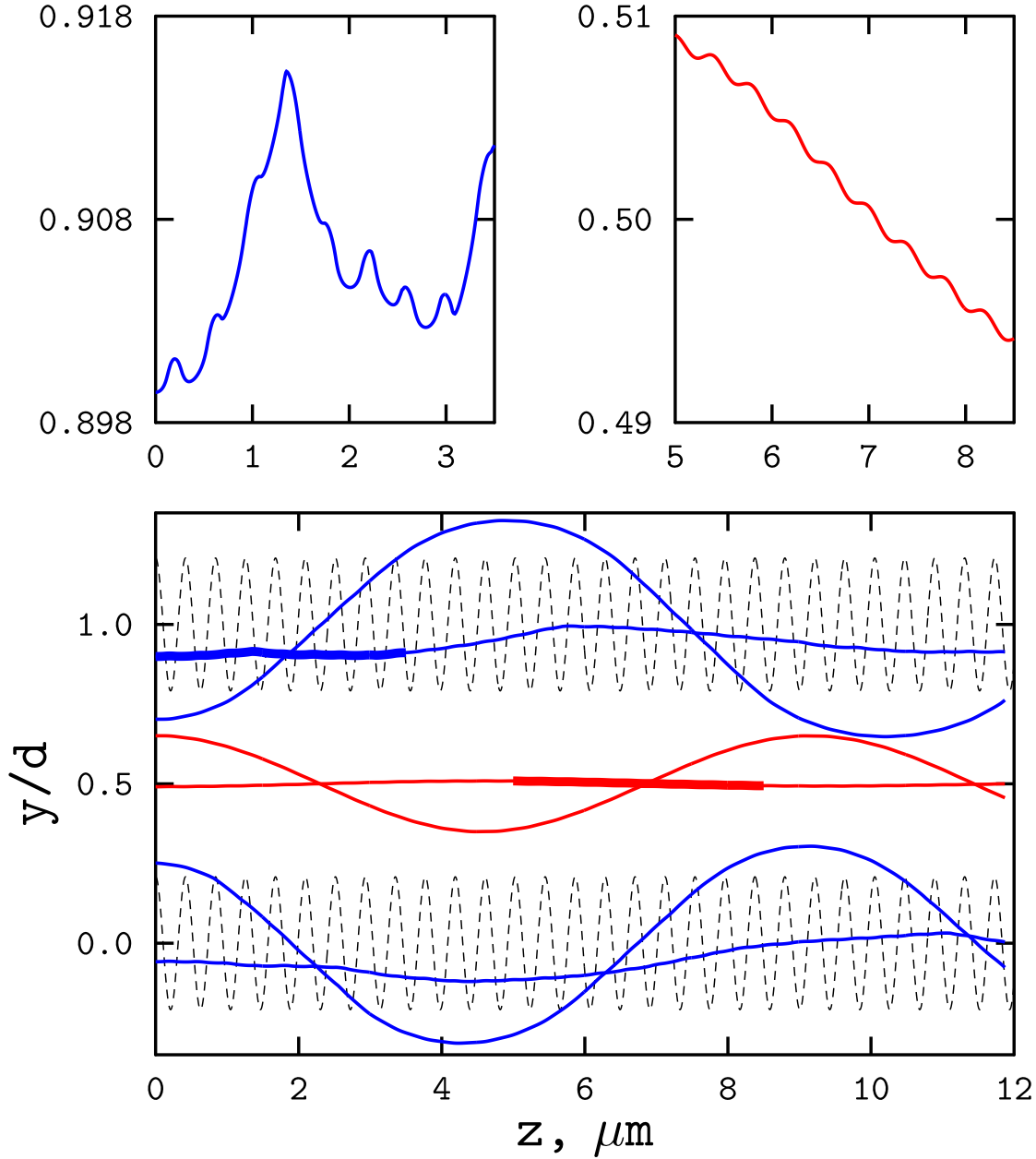}
%\vspace{-5mm}
\caption{(color online). Channeling of the $10$~GeV projectiles through a $12~\mu$m thick 
Si undulator with $\lamu = 400$~nm and $a = 0.4$~{\AA}. The trajectories and the crystalline 
profiles are shown by the same types of lines as in Fig.~\ref{trajectories_855}. 
The short-period undulator oscillations are prominent in the trajectories with small mean 
amplitude of the transverse oscillations - see the separate plots corresponding to the segments 
displayed by thicker lines.}  
\label{trajectories}
%\vspace{-4mm}
\end{figure}

Next we have studied the channeling and radiation for the multi-CeV particles. 
Having analyzed the simulated trajectories, we can confirm the channeling to indeed occur 
for the $10$~GeV projectiles, both for the electrons and positrons, in the short-period 
small-amplitude CUs. The variety of typical channeling trajectories appears to be distributed 
between two distinct trajectory types different in mean amplitude of the transverse oscillations. 
The latter types are shown in Figure~\ref{trajectories} and are selected from the simulations for 
the CU with $\lamu = 400$~nm and $a = 0.4$~{\AA}. Also shown in the figure are the bent profiles for 
the (110) planes in Si. The trajectories and profiles are plotted in terms of the transverse 
coordinate $y$ (in units of the inter-planner distance $d = 1.92$~{\AA} for the straight Si crystal) 
and the longitudinal coordinate $z$ (in $\mu$m). We remark that our simulations are the 3D ones, and 
the 2D plots are provided for better visualization of the dynamics of channeling. 
 
\begin{figure}[th]
%\hspace{-6mm}
\includegraphics[width=\columnwidth]{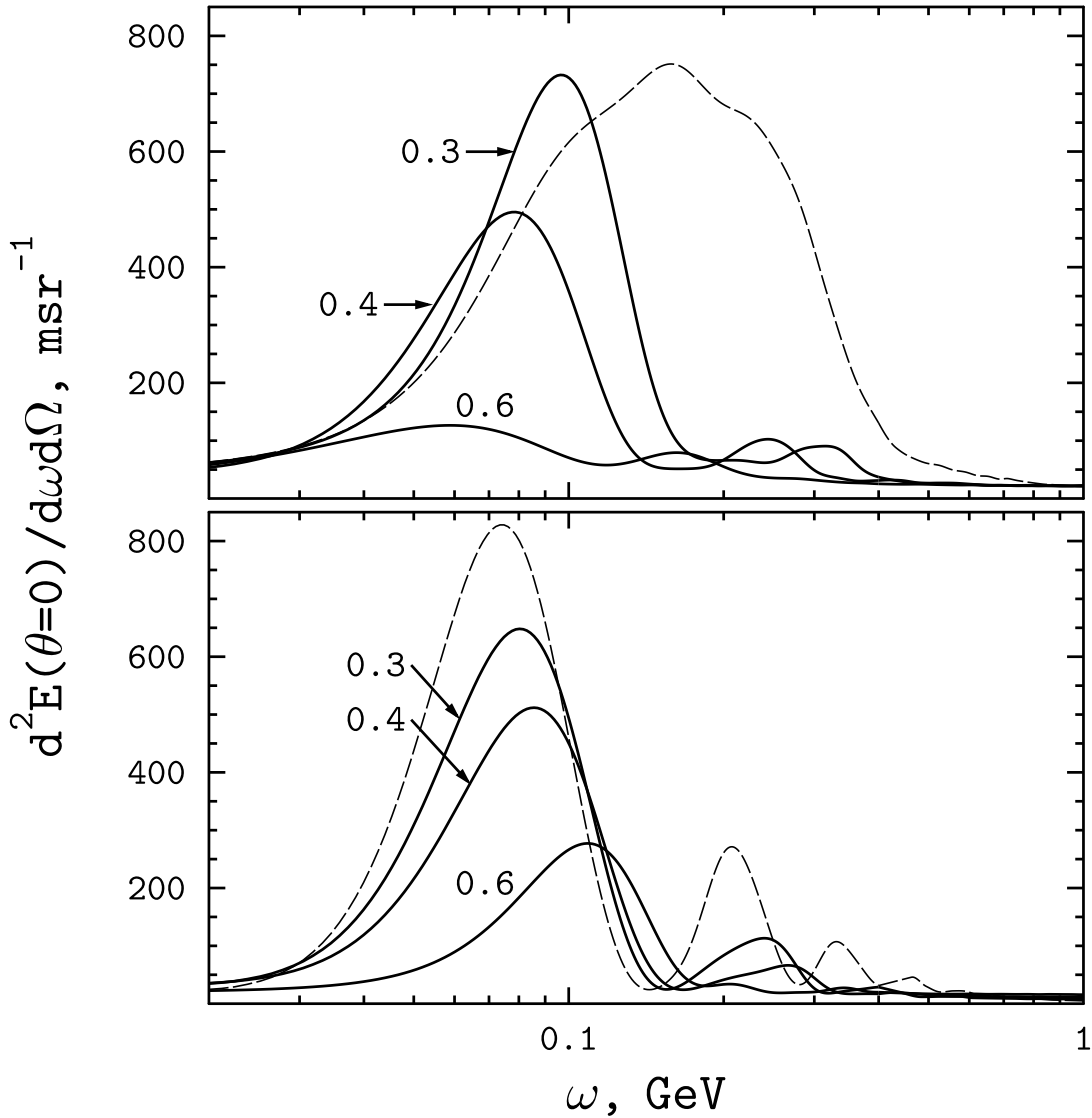}
%\vspace{-10mm}
\caption{Impact of the bending on the channeling radiation by electrons (upper plot) and 
positrons (bottom plot). Thin dashed curves are the spectra for the planar channeling in a 
straight $12~\mu$m Si(110). Thick solid curves are the spectra for the cosine bending profiles of the 
crystal, with the period $400$~nm and different amplitudes (the amplitude values in {\AA} are indicated 
near the curves).}  
%\vspace{-3mm}
\label{e_p_400_03_04_06}
\end{figure}    

The short-period bent crystalline profiles in Figure~\ref{trajectories} define the channels guiding the 
projectiles through the crystal. In the channeling mode, the electron and positron trajectories pass 
through and in between the bent profiles, respectively. One type of the trajectories studied in the 
figure are the long-period large-amplitude trajectories. These exhibit two large oscillations along the 
$12~\mu$m pass through the crystal. For the electrons, the oscillation amplitudes slightly exceed the 
bending amplitude $a$, whereas for the positrons the oscillations get within nearly all the space free 
from the atoms in between the bent crystalline planes. Another kind of the trajectories are the 
short-period small-amplitude ones. On the scale of bending amplitude, the corresponding motions 
are rather close to a translational motion of the free projectiles, though for the electrons in a 
valuably less extent than for the positrons. Yet on the smaller transverse scale, these trajectories 
display the oscillations that obviously result from the bent crystalline profiles (see the parts 
of the trajectories displayed in two separate plots in the figure). The maximum number of the 
oscillations in a trajectory is given by the ratio $12~\mu\mbox{m}/\lamu$, i.e. is 
$30$ for the CU with $\lamu = 400$~nm studied in Figure~\ref{trajectories}. 

\begin{figure*}[th]
\centering
\begin{tabular}{cc}
\hspace{-2mm}
\includegraphics[width=0.5\textwidth]{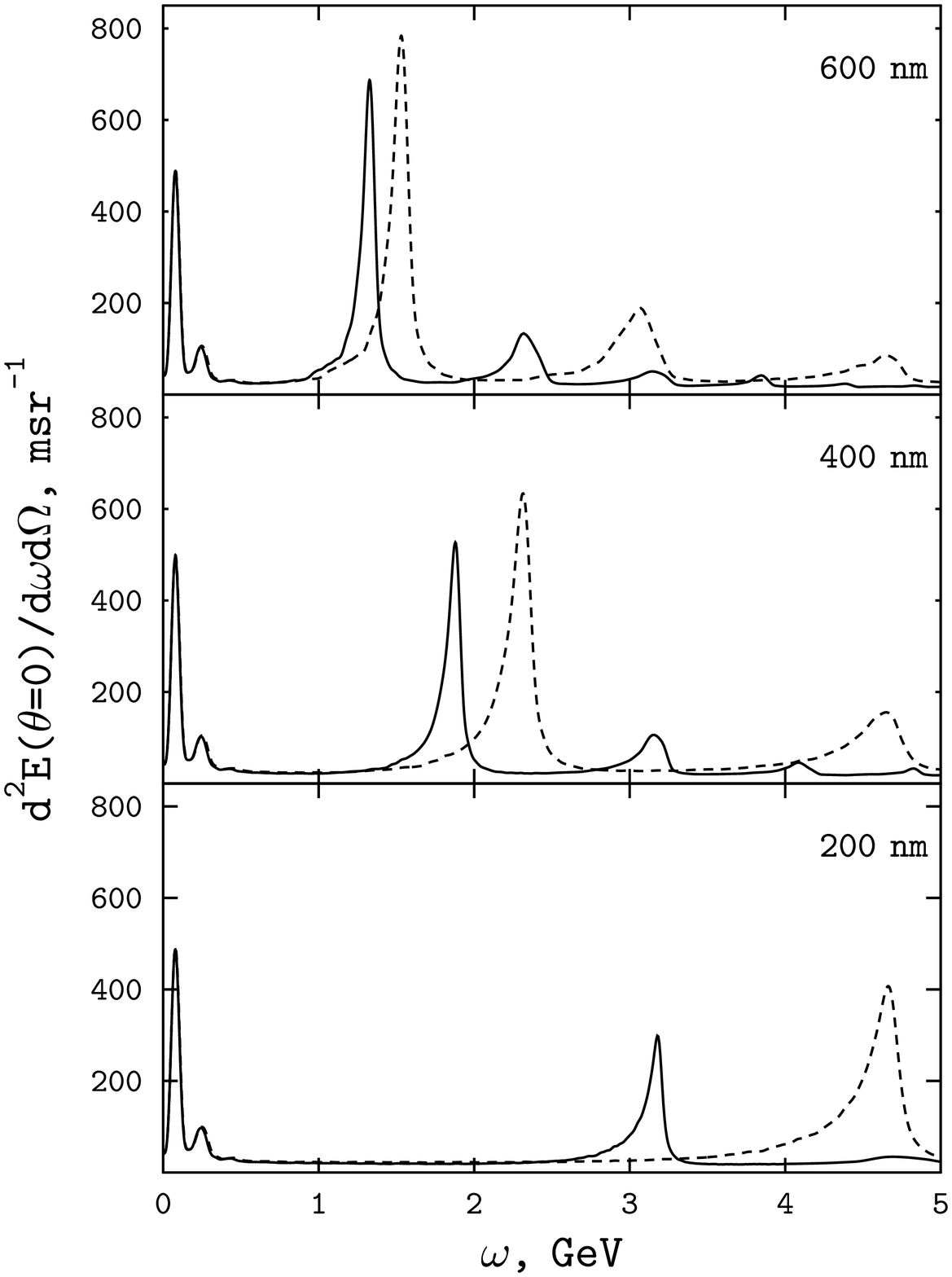} &
\hspace{-25mm}
\includegraphics[width=0.5\textwidth]{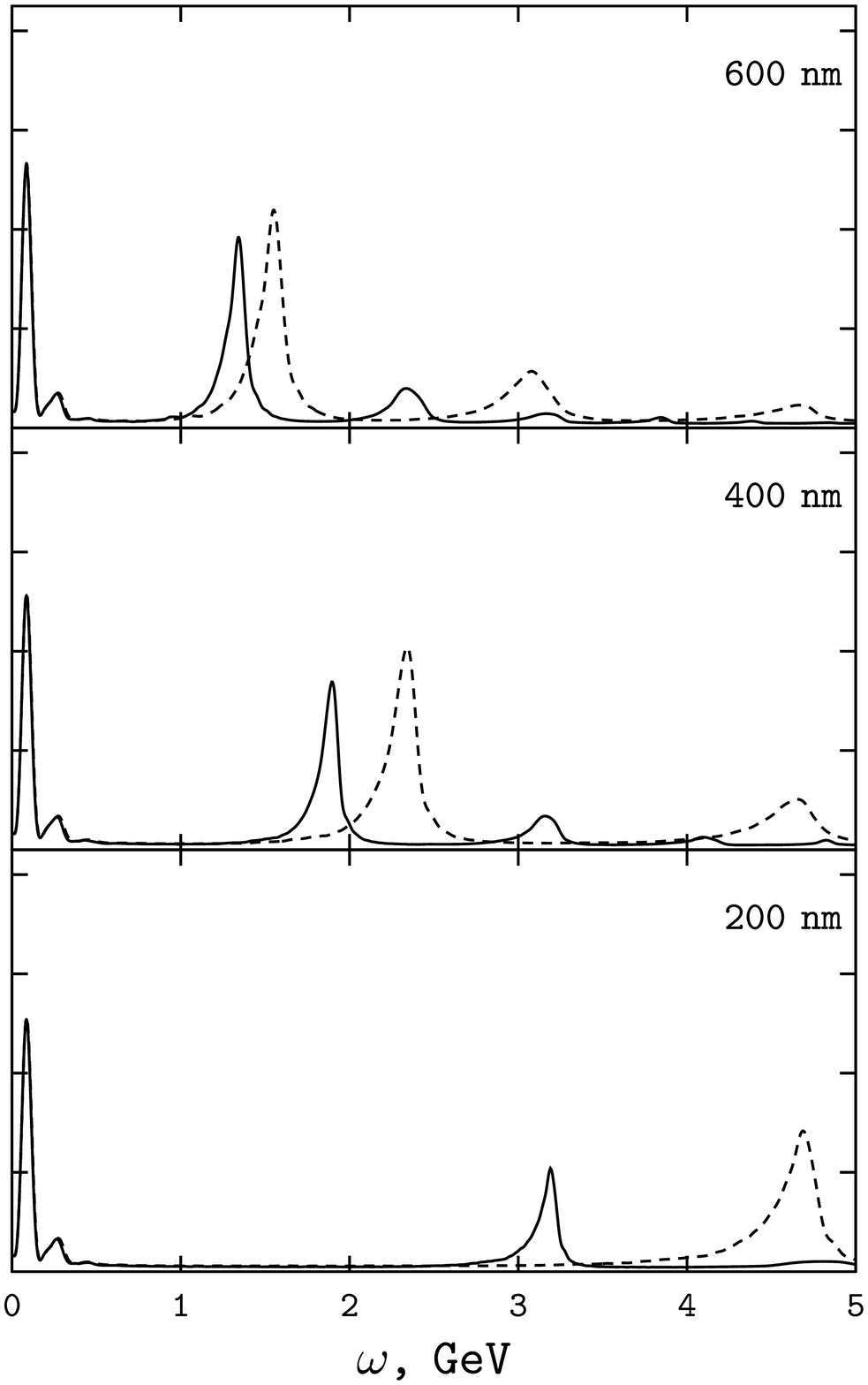}
\end{tabular}
\caption{Radiation spectra produced by the projectiles of the energy $\E = 10$~GeV propagating through 
a $12~\mu$m thick Si undulator. The crystalline channels have cosine profiles with the amplitude 
$0.4~\mbox{\AA}$. Left and right panels show the spectra by electrons and positrons, respectively, 
for different values of the bending period indicated in the upper right corners of the plots. 
The dashed-line curves are the spectra computed according to the classical formalism (i.e. assuming 
$\omega/\E \to 0$ and neglecting thereby the quantum recoil).}  
\label{spectra_10_04_200_400_600}
%\vspace{-3mm}
\end{figure*}    

To this end, we focus on the properties of CU radiation by the $10$~GeV projectiles. Our concern are 
the spectral intensities $\mbox{d}^2E(\theta=0)/\mbox{d}\omega\mbox{d}\Omega$ calculated as the 
energy $\mbox{d}^2E$ emitted in the direction of the mean velocity of channeling particles 
($\theta = 0$) per photon energy $\mbox{d}\omega$ and per solid angle $\mbox{d}\Omega$.

Two kinds of oscillations in the trajectories discussed above contribute differently to the radiation 
spectra. The long-period oscillations determine the channeling radiation with the peak 
intensities in the soft energy range. Thought this radiation is similar to the radiation from straight 
crystal, it is influenced by the bending. In particular, increasing the bending amplitude suppresses the 
channeling radiation. The latter effect is demonstrated in Figure~\ref{e_p_400_03_04_06} and is related 
to decreasing portion of the long-period large-amplitude trajectories in CUs with larger bending 
amplitudes.   

The short-period trajectories contribute to the undulator radiation from the CU. The corresponding 
spectral lines appear at the gamma-ray energies, as demonstrated in 
Figure~\ref{spectra_10_04_200_400_600}. The spectra in this figure are computed for CUs with the 
amplitude $a=0.4$~{\AA} and three periods, $\lamu = 600$~nm, $\lamu = 400$~nm and $\lamu = 200$~nm, 
respectively. Shown by the solid lines are the spectra computed with account for the quantum recoil, 
whereas the dashed lines are the results of neglecting the recoil in calculations. An immediate impact 
of the figure are the striking differences in the positions of the undulator peaks computed with and 
without accounting for the recoil. The latter is not a surprise as including the recoil yields the 
fundamental undulator peaks at the energies $1-3$~GeV far not negligible compared to the 
beam energy of $10$~GeV. These peaks appear at the softer energies, as compared to the peaks 
computed by employing a pure classical formalism. We also notice that the 
recoil yields the fundamental and higher undulator harmonics to be non-equidistant in spectral 
locations, in contrast to the equidistant peaks computed within the classical approximation. 
In contrast to the undulator peaks, the channeling peaks emerge at the energies $0.1-0.3$~GeV much 
smaller than the beam energy. Therefore the channeling parts of the spectra are not influenced by the 
recoil in any noticeable way. 

\begin{table}[th]
%\vspace{3mm}
\renewcommand*{\arraystretch}{1.2}
\begin{tabular}{|c|c|c|cc|c|cc|c|}
\cline{4-9}
\multicolumn{3}{c|}{}                  & \multicolumn{3}{c|}{electrons}    & \multicolumn{3}{c|}{positrons}    \\
\hline
$\lamu$ & $\Omega_u$ & $2\gamma^2\Omu$ & $\omega'_1$ & $\omega_1$ & $K$    & $\omega'_1$ & $\omega_1$ & $K$    \\
nm      & eV         & GeV             & \multicolumn{2}{c|}{GeV} &        & \multicolumn{2}{c|}{GeV} &        \\
\hline
$600$   & $2.06$     & $1.58$          & $1.53$      & $1.33$     & $0.26$ & $1.55$      & $1.34$     & $0.20$ \\
\hline
$400$   & $3.09$     & $2.38$          & $2.31$      & $1.88$     & $0.23$ & $2.34$      & $1.90$     & $0.18$ \\
\hline
$200$   & $6.18$     & $4.75$          & $4.66$      & $3.18$     & $0.20$ & $4.69$      & $3.19$     & $0.16$ \\
\hline
\end{tabular}
\caption{Undulator frequencies for the motion and the spectral lines for the $10$~GeV projectiles in CU with 
the amplitude $a=0.4$~{\AA} and different periods $\lamu$.}
\label{undulator_energies}
\end{table}

In Table~\ref{undulator_energies} we provide the central frequencies for the undulator peaks. 
The frequencies $\omega_1$ and $\omega'_1$ are determined from the numerical simulations of the 
spectra including and neglecting the quantum recoil, respectively, and appear to be in a decent 
agreement with the theoretical result~(\ref{peak_positions}). Also included in the table are the 
values for the undulator parameter $K$ which are deduced from comparing $\omega_1$ and $\omega'_1$ 
with the values $2\gamma^2\Omu$. For all periods of CU studied, the values of $K$ drop far below the 
value of $1$ implying thereby that the bent crystals work indeed as the undulators for the channeling 
multi-GeV projectiles. With decreasing bending period, the values of $K$ decrease as well being slightly 
smaller for the positrons than for the electrons.

As the short-period small-amplitude bent Si crystals are being manufactured and the high-quality beams 
of electrons and positrons with multi-GeV energies are on the present experimental disposal~\cite{Ugg}, 
it is demanding to bring these ingredients together for producing and manipulating the monochromatic 
gamma-ray radiation. 
Theoretical support of the experiments has to include the effects of the quantum recoil in order 
to deliver reliable predictions on the radiation spectra. The experimental measurements can also 
provide the values for the undulator parameter which are valuable for theory of the channeling 
phenomena.
 
\ack
Financial support due to the European CUTE Project is gratefully acknowledged. 
We are grateful to H. Backe, W. Lauth and G. Sushko for fruitful discussions, and 
acknowledge the particular contribution of G. Sushko in developing the MBN Explorer 
and setting up various aspects of the channeling simulations.


\section*{References}

\begin{thebibliography}{99}

\bibitem{ChannelingBook2013} 
Korol A V, Solov'yov A V and Greiner W 2013 
{\em Channeling and Radiation in Periodically Bent Crystals} (Springer-Verlag Berlin Heidelberg)

\bibitem{NewPaper_2013}
Sushko G B, Bezchastnov V G, Solov'yov I A, Korol A V, Greiner W and Solov'yov A V 2013 
{\em J. Comp. Phys.} \textbf{252} 404 

\bibitem{JPCS_2013a} 
Sushko G B, Korol A V, Greiner W and Solov’yov A V 2013 
{\em J. Phys. Conf. Ser.} \textbf{438} 012018

\bibitem{JPCS_2013b}
Sushko G B, Bezchastnov V G, Korol A V, Greiner W, Solov’yov A V, Polozkov R G and Ivanov V K 2013 
{\em J. Phys. Conf. Ser.} \textbf{438} 012019

\bibitem{K_PRL_2013}
Kostyuk A 2013 
{\em Phys. Rev. Lett.} \textbf{110} 115503 

\bibitem{Mainz_PRL_2014}
Mazzolari A, Bagli E, Bandiera L, Guidi V, Backe H, Lauth W, Tikhomirov V, Berra A, Lietti D, 
Prest M, Vallazza E and Salvador D 2014 
{\em Phys. Rev. Lett.} \textbf{112} 135503 

\bibitem{Jackson}
Jackson J D 1999
{\em Classical Electrodynamics} 3rd edn (New York: Wiley)

\bibitem{KSG1998} 
Korol A V, Solov'yov A V and Greiner W 1998  
{\em J. Phys. G} \textbf{24} L45 

\bibitem{KSG_review_1999} 
Korol A V, Solov'yov A V and Greiner W 1999 
{em Int. J. Mod. Phys. E} \textbf{8} 49 

\bibitem{Ugg} 
Uggerh{\o}j U I 2008 
{\em Recentr crystalline undulator experimental developments} talk at the PECU Meeting, 
Frankfurt am Main, Germany, 18 Jan. 2008 (unpublished) 

\bibitem{MBN_ExplorerPaper} 
Solov'yov I A, Yakubovich A V, Nikolaev P V, Volkovets I and Solov'yov A V 2012 
{\em J. Comp. Chem.} \textbf{33} 2412 

\bibitem{MBN_ExplorerSite} http://www.mbnexplorer.com/

\bibitem{Baier_1988} 
Baier V N, Katkov V N and Strakhovenko V M 1988  
{\em Electromagnetic Processes at High Energies in Oriented Single Crystals}, 
(Singapore: World Scientific) 
  
\end{thebibliography}
\end{document}